\documentclass[showpacs,twocolumn,preprintnumbers,amsmath,amssymb,amsfonts,prl,floatfix,superscriptaddress]{revtex4}

\usepackage{graphicx, color,rotating} 
\usepackage{dcolumn} 
\usepackage{bm} 
\usepackage{comment}
\usepackage{multirow}
\newcommand{\PP}{{\textrm{\tiny P}}}
\newcommand{\AP}{{\textrm{\tiny A$\!$P}}}

\begin{document}

\title{Nonlocal conductance via overlapping Andreev bound states
in ferromagnet-superconductor heterostructures}

\author{Georgo~Metalidis}
\affiliation{Institut f\"ur Theoretische Festk\"orperphysik
and DFG-Center for Functional Nanostructures,
Karlsruhe Institute of Technology (KIT), D-76128 Karlsruhe, Germany}
\author{Matthias~Eschrig}
\affiliation{Institut f\"ur Theoretische Festk\"orperphysik
and DFG-Center for Functional Nanostructures,
Karlsruhe Institute of Technology (KIT), D-76128 Karlsruhe, Germany}
\affiliation{Fachbereich Physik, Universit\"at Konstanz, D-78457 Konstanz, Germany}
\author{Roland~Grein}
\affiliation{Institut f\"ur Theoretische Festk\"orperphysik
and DFG-Center for Functional Nanostructures,
Karlsruhe Institute of Technology (KIT), D-76128 Karlsruhe, Germany}
\author{Gerd~Sch\"on}
\affiliation{Institut f\"ur Theoretische Festk\"orperphysik
and DFG-Center for Functional Nanostructures,
Karlsruhe Institute of Technology (KIT), D-76128 Karlsruhe, Germany}

\date{\today}

\begin{abstract}
In a setup where two ferromagnetic electrodes are attached to a superconductor, Andreev bound states are induced at both ferromagnet/superconductor (FM/SC) interfaces. We study how these states propagate through the SC and interact with each other. As a result of this interaction, the energetic positions of the Andreev states
are not anymore determined solely
by the magnetic properties of a single interface, but also depend on the interface distance and the relative magnetization orientation of the FM contacts.
We discuss how
these bound states show up as distinct peaks in the nonlocal conductance signal, and lead to marked asymmetries with respect to the applied voltage.
We relate our results to nonlocal crossed Andreev and elastic co-tunneling processes.
\end{abstract}


\maketitle

\newcommand{\gammat}{\tilde{\gamma}}
\newcommand{\Gammat}{\tilde{\Gamma}}
\newcommand{\x}{x}
\newcommand{\xt}{\tilde{x}}
\newcommand{\X}{X}
\newcommand{\Xt}{\tilde{X}}
\renewcommand{\i}{\mathrm{i}}
\newcommand{\Deltat}{\tilde{\Delta}}
\renewcommand{\S}{S}
\newcommand{\St}{\tilde{S}}
\newcommand{\ura}[1]{\underrightarrow{#1}}
\newcommand{\ula}[1]{\underleftarrow{#1}}
\newcommand{\sgn}[1]{\mathrm{sign}({#1})}

\newcommand{\eps}{\varepsilon}
\newcommand{\gr}{\gamma^R}
\newcommand{\ga}{\gamma^A}
\newcommand{\grt}{\tilde{\gamma}^R}
\newcommand{\gat}{\tilde{\gamma}^A}
\newcommand{\gra}{\gamma^{R,A}}
\newcommand{\grat}{\tilde{\gamma}^{R,A}}

The presence of Andreev bound states (ABS) at energies below the superconducting gap plays a prominent role in transport through heterostructures involving superconductors, and forms a topic of continuing and long-standing interest. Such states can arise, e.g., due to multiple Andreev reflections at both interfaces of a SC-insulator-SC junction, and are directly related to the Josephson current through the junction~\cite{Deutscher2005}. Furthermore, the importance of a magnetically active interface between a SC and FM has been pointed out. Scattering on such a surface leads to  spin-mixing and the creation of subgap
ABS
at the SC/FM interface~\cite{DeWeert1985,Fogelstrom2000}. Such states influence the properties of, e.g., single SC/FM tunnel junctions~\cite{Cottet2008, Grein2010} and SC/FM/SC Josephson junctions~\cite{Fogelstrom2000,Eschrig2003}.

Andreev reflection processes assume a nonlocal (NL) form when two metallic leads are connected to the SC: the incoming quasiparticle in one lead can be  reflected as a quasihole in the other lead~\cite{Deutscher2000}, giving rise to a negative NL conductance. This so-called crossed Andreev reflection (CAR) competes with elastic co-tunneling (EC) (and, at high temperatures, charge imbalance
effects~\cite{Tinkham1972}),
in which the quasiparticle is transferred from one lead to the other via a virtual subgap state in the SC and gives a positive contribution to the NL conductance~\cite{Beckmann2004}. CAR and EC processes cancel each other for tunnel contacts~\cite{Falci2001}. This cancelation is lifted at higher orders in the transmission~\cite{Melin2004}, for FM contacts~\cite{Falci2001, Melin2004, Yamashita2003, Kalenkov2007}, or in the presence of
interactions~\cite{Yeyati2007}.
Disorder effects have been
addressed in Refs.~\cite{Feinberg2003}.
Even though solid theoretical progress has been made, the role of Andreev bound states in the nonlocal conductance has not been elucidated.

\begin{figure}[b]
\includegraphics[width = 0.88 \columnwidth]{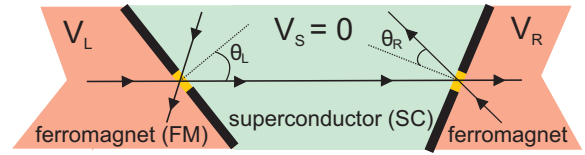}
\caption{(Color online) Sketch of the considered FM/SC/FM setup with
two tunneling point contacts.
A nonlocal scattering event which involves both contacts
is indicated.
\label{fig:setup}}
\end{figure}

In this Letter, we consider the non-local setup shown in Fig.~\ref{fig:setup}, where two FM point contacts are attached to a ballistic SC region. At each of the two SC/FM contacts subgap ABS form, which show up in the Andreev spectrum~\cite{Fogelstrom2000,Grein2010}.
They propagate on a coherence length scale $\xi $ through the SC. Interestingly,
these states interact with each other so that their energetic position is not only dependent on the magnetic properties of the separate interfaces, but also on the distance between the interfaces and their relative magnetization orientation. We study the profound influence of this ABS interaction on the NL conductance through the device when voltages $V_{\rm L}$ and $V_{\rm R}$ are applied across the contacts. For identical contact parameters,
we find that for parallel (P) magnetization, the NL conductance $\partial I_{\rm R} / \partial V_{\rm L}$ is asymmetric in $V_{\rm L}$, whereas it stays symmetric in the antiparallel (AP) configuration. Such (a)symmetries will be explained with an intuitive picture based on the ABS positions in the system.

Transport properties of heterostructures involving SCs are conveniently described using quasiclassical Green's functions (QC GFs), $\hat{g}(\vec{v}_{\rm{F}} , \vec{R}, \eps)$, which depend on the Fermi velocity $\vec{v}_{\rm{F}}$, the spatial coordinate $\vec{R}$ and the quasiparticle energy $\eps$, and obey the Eilenberger equation
\begin{equation} \label{eq:Eilenberger}
\i \hbar \vec{v}_{\rm{F}} \cdot \nabla_{\vec{R}}\hat{g} + [\eps\hat{\tau}_3  - \hat{\Delta}, \hat{g}]=\hat{0},
\end{equation}
subject to the normalization condition $\hat{g}^2=-\pi^2 $. The ''hat'' refers to the $2\times2$ matrix structure of the propagator in particle-hole (Nambu) space, and $\hat{\Delta}$ is the SC order parameter. Within our approach,
the exchange energy is incorporated by different Fermi velocities $\vec{v}_{\rm{F} \sigma}$ and momenta $\vec{p}_{\rm{F} \sigma}$ in the spin bands \cite{Grein2009,Eschrig2009}. Thus, all elements of the SC propagators, $\hat{g}_{\text{SC}}$, are $2\times2$ spin-matrices, while the elements of the FM propagators, $\hat{g}_{\sigma}$, are scalars.

A SC/FM interface enters the QC theory as boundary conditions relating outgoing and incoming propagators~\cite{Eschrig2009}. They are expressed in terms of the normal state scattering matrix $S$ of the interface.
We assume that the transmission ($t^{\alpha \beta}$) and reflection coefficients ($r^{\beta }$) of the $S$-matrix are spin-diagonal with respect to the magnetization of the FM, with diagonal elements $t_\sigma \exp(\i \vartheta^{\alpha \beta}_\sigma ) $ and $r_\sigma \exp(\i \vartheta^\beta_{\sigma })$ where $\sigma \in \{\uparrow,\downarrow\}$ and $\alpha,\beta \in \{{\rm SC, \rm FM}\}$.
\begin{figure}
\includegraphics[width = 1.00\columnwidth]{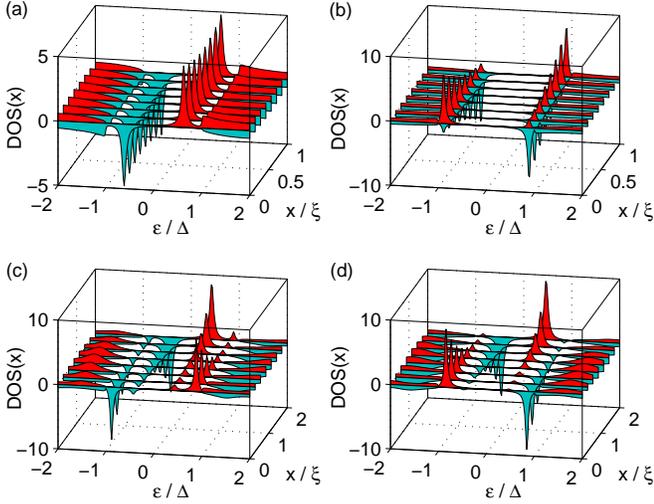}
\caption{(Color online) Spin-up (positive axis) and spin-down (negative axis) DOS at different positions $x$ for parallel
(left column)
and antiparallel
magnetization
(right column).
(a) and (b): $\rho = 0.9$, $L=\xi$,
$\vartheta_{\rm R} = \pm \vartheta_{\rm L} = 0.4\pi$.
(c) and (d): $\rho=0.9$, $L=2\xi $,
$\vartheta_{\rm L} \neq \vartheta_{\rm R}$, with
$\vartheta_{\rm L} = \pm0.4\pi$, $\vartheta_{\rm R} = 0.7\pi$. \label{fig:LDOSvsx}}
\end{figure}

First we calculate the local density of states (DOS)
at points
along the line connecting both FM contacts and investigate the ABS interaction.
We concentrate on the contribution arising
from the trajectory that connects the two contacts (see Fig.~\ref{fig:setup}).
Denoting $x$ as
the distance from the left electrode, we obtain from the solution of
Eq.~\eqref{eq:Eilenberger}
\begin{equation} \label{eq:LDOSspinresolved}
\frac{N_{\sigma}(x,\eps) }{N_{\rm F}}= \frac{1}{2}
\textrm{Re} \left[ (1 - \Gamma_{{\rm L}\sigma} \Gammat_{{\rm R}\bar{\sigma}})/(1 + \Gamma_{{\rm L}\sigma} \Gammat_{{\rm R}\bar{\sigma}})\right],
\end{equation}
where $N_{\rm F}$ is the normal state DOS in the SC, and with the notation
$\bar{\sigma} = \uparrow (\downarrow)$ when $\sigma = \downarrow (\uparrow)$. For subgap energies, the $\Gamma_{i\sigma}$ with $i\in \{\rm L,R\}$
originating from the left (L) and right (R) contact,
are given by \cite{Kalenkov2007,Eschrig2009}
\begin{eqnarray}
\Gamma_{i\sigma}(x,\varepsilon) &=& \left[\Omega \gamma_0
s_{i\sigma}
+ \i \left( \gamma_0 \varepsilon
s_{i\sigma}
+ \Delta \right) \tanh(\Omega x_{i}/ \Delta\xi )\right] \nonumber \\
&\times& \left[\Omega - \i \left( \gamma_0 \Delta
s_{i\sigma}
+ \varepsilon \right) \tanh(\Omega x_{i}/\Delta \xi) \right]^{-1}, \label{eq:GammaLR}
\end{eqnarray}
and the $\Gammat_{i\sigma}$ result from $\Gammat_{i\sigma}(x,\eps) = \Gamma_{i\sigma}^\ast(x,-\eps)$.
Here,
$\Omega = \sqrt{\Delta^2 - \varepsilon^2}$,
$\gamma_0 = (\i \Omega - \varepsilon)/\Delta$,
$s_{i\uparrow } = \rho_{i} \mathrm{e}^{\i \vartheta_i} $,
$s_{i\downarrow} = \rho_{i} \mathrm{e}^{-\i \vartheta_i} $,
$x_{\rm L} = x$, $x_{\rm R} = L-x$ (where $L$ is the
distance between the contacts),
and $\xi = \hbar v^\textrm{\tiny SC}_{\rm F}/\Delta $.
Interface properties enter the DOS only via the product $\rho = r_\uparrow r_\downarrow$ and the spin-mixing angle $\vartheta \equiv \vartheta^\textrm{\tiny SC}_\uparrow - \vartheta^\textrm{\tiny SC}_\downarrow$
at left and right interface.
In the following, we assume for simplicity $\rho = \rho_{\rm L} = \rho_{\rm R}$, but allow for differing $\vartheta_{\rm L}$ and $\vartheta_{\rm R}$. In particular, $\vartheta_{\rm R} = (-) \vartheta_{\rm L} \equiv \vartheta$ when speaking of P (AP) magnetization.

In Fig.~\ref{fig:LDOSvsx}, the DOS is shown at different positions $x$ between the interfaces. ABS appear as distinct spin-resolved subgap peaks. In general, for a positive spin-mixing angle, a spin-up state appears at positive energy. For negative spin-mixing angles, spin-up and spin-down peak positions are interchanged, as expected from Eq.~(\ref{eq:GammaLR}). Fig.~\ref{fig:LDOSvsx} reflects this behavior. For the P case $\vartheta_{\rm L} = \vartheta_{\rm R} > 0$, the peaks appear at identical energies at the left ($x=0$) and right ($x=L$) interface.
For AP magnetization with $\vartheta_{\rm L} = -\vartheta_{\rm R} < 0$ [Fig.~\ref{fig:LDOSvsx}(b)] the spin-up (down) state is created at a negative (positive) energy at the left interface, where $\vartheta_{\rm L} < 0$, but at a positive (negative) energy for the right interface, where $\vartheta_{\rm R} > 0$. The DOS peak height decreases upon propagation through the SC. For AP magnetization, this leads to two spin-up (and down) states at each interface: the smaller one is created at the opposite interface and is attenuated upon propagation through the SC. This attenuation is less pronounced in the P case, where it is countered by the propagating state from the other interface. If $|\vartheta_{\rm L}| \neq |\vartheta_{\rm R}|$, ABS have different energies at left and right interface, and their spin-character depends on the sign of the spin-mixing angles [compare Figs.~\ref{fig:LDOSvsx}(c) and~(d)].

These states correspond to poles in Eq.~(\ref{eq:LDOSspinresolved}).
From equating $1 + \Gamma_{{\rm L}\sigma} \Gammat_{{\rm R}\bar{\sigma}}=0$ we obtain in the tunneling
limit ($\rho_{\rm L} = \rho_{\rm R} = 1$)
\begin{equation} \label{eq:boundstates}
\cos(\psi+\frac{\vartheta_{\rm L}}{2}) \cos(\psi+\frac{\vartheta_{\rm R}}{2}) = \mathrm{e}^{-\frac{2L}{\xi}\cos\psi }\sin(\frac{\vartheta_{\rm L}}{2})\sin(\frac{\vartheta_{\rm R}}{2})
\end{equation}
for the spin-up ABS energies at the right interface. Spin-down states are found from $N_\sigma(x,\eps) = N_{\bar{\sigma}}(x,-\eps)$. The energy dependence in Eq.~(\ref{eq:boundstates}) is contained in the variable $\psi = \arcsin\left(\varepsilon/\Delta\right)$. ABS positions thus not only depend on the spin-mixing angle, but also on the interface distance $L$ and the magnetization configuration of the FMs.
For lower $\rho$, i.e., higher interface transmissions, the DOS peaks are broadened into
resonances.
\begin{figure}
\includegraphics[width = \columnwidth]{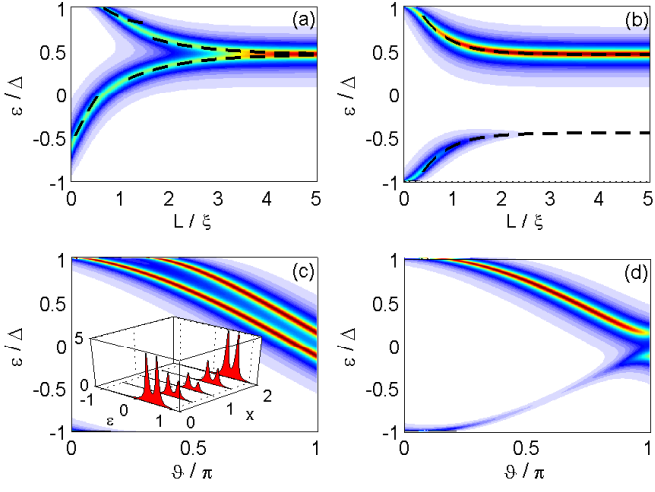}
\caption{(Color online) Spin-up DOS at the right interface vs $L$~(upper panel) and $\vartheta$~(lower panel), both for P (a,c) and AP (b,d) magnetization. Dashed lines are analytical approximations (see text). Inset: propagation of the split ABS between the interfaces. Parameters in Figs~(a,b): $\rho = 0.9$, $\vartheta = 0.7 \pi$. In Figs.~(c,d): $\rho=0.9$, $L=2.0 \xi$. Inset: $\vartheta = 0.7 \pi$, $L=2.0 \xi$.\label{fig:LDOSvsLengthandtheta}}
\end{figure}

In Fig.~\ref{fig:LDOSvsLengthandtheta}, the spin-up DOS at the right interface ($x=L$) is plotted as a function of the interface distance $L$ and the spin-mixing angle $\vartheta$. For the P configuration, one obtains a single spin-up ABS at $\varepsilon_\infty^{\rm P} = \Delta \cos(\frac{\vartheta}{2})$ in the limit $L/\xi \rightarrow \infty$. This state splits in two as soon as the interface distance $L$ is decreased. This can be interpreted as follows: the ABSs originating at both interfaces [see Fig.~\ref{fig:LDOSvsx}(a)], having the same energy in the limit $L/\xi \rightarrow\infty$, repel each other as the interfaces are brought closer. To first order, the energy correction to $\varepsilon_\infty^{\rm P}$ is given by $\delta \varepsilon^{\rm P} = \Delta \sin^2 (\frac{\vartheta}{2}) e^{-\frac{L}{\xi}\sin (\frac{\vartheta}{2}) }$, assuming $\delta \varepsilon \ll |\Delta - \varepsilon_\infty^{\rm P}|$. So the repulsion strength increases with $\vartheta$. When the interfaces are close enough, the higher energy state is pushed out of the SC gap; close to the gap, its energy can be approximated by $\Delta - 2 \Delta [(L/\xi)-\cot(\frac{\vartheta}{2})]^2 / [1+(L/\xi)^2]^2$. The critical length at which the state enters the continuum is thus given by $L=\xi \cot(\frac{\vartheta}{2})$. At these small lengths, the lower-energy state goes like
$\varepsilon_0^{\rm P} +
2\Delta \sin^2 (\frac{\vartheta}{2}) (1-e^{-\frac{L}{\xi}\sin\vartheta })$
with $\varepsilon_0^{\rm P} = \Delta \cos(\vartheta)$ its energy for $L=0$.
When $\vartheta >\frac{\pi }{ 2 }$, a zero bias state exists for
$L=\xi \ln |\tan (\frac{\vartheta }{2})|$.
The propagation of ABS split peaks through the SC is sketched in the inset of Fig.~\ref{fig:LDOSvsLengthandtheta}(c); the weight of the higher energy state is
smaller than that of the lower one and
decays faster.

In the AP case, there are two ABS at energies $\pm\varepsilon_\infty^{\rm AP} = \pm\Delta \cos(\frac{\vartheta}{2})$ in the limit $L/\xi \rightarrow \infty$. In terms of Fig.~\ref{fig:LDOSvsx}(b), the negative energy state corresponds to the ABS that has
propagated from the other interface. Its weight is therefore reduced (it is zero for $L/\xi \rightarrow \infty$). As the interfaces are brought closer to each other, this state gains in weight and both ABS repel each other. To first order, their energies are $\pm\varepsilon_\infty^{\rm AP} \pm \frac{1}{2}\Delta \tan\frac{\vartheta}{2} e^{-\frac{2L}{\xi}\sin(\frac{\vartheta}{2})}$. They both enter the continuum at $L=0$.

The influence of the spin-mixing angle $\vartheta$ is depicted in Figs.~\ref{fig:LDOSvsLengthandtheta}(c) and ~(d). Increasing $\vartheta$ moves all states deeper inside the gap. For the P case, the second ABS only appears for values $\vartheta > 2 \textrm{arctan} (L/\xi)$. In the AP case, the negative-energy state is weakened as it must propagate through the SC to reach the right interface.

These Andreev states show up in the transport properties of FM/SC/FM heterostructures, in particular in the NL conductance $G_{\rm NL}$. Assuming a constant current density $j$ over the contact surface area, the NL current $I_{\rm R}$ in the right contact is:
\begin{equation}
I_{\rm R} = \frac{{\cal S}_{\rm L} {\cal S}_{\rm R} \cos(\theta_{\rm L}) \cos(\theta_{\rm R})}{L^2} \, j,
\end{equation}
where ${\cal S}_{\rm L(R)}$ is the area of the left (right) contact,
$L$ is again the distance between the contacts,
and $\theta_{\rm L(R)}$ is the impact angle at the left (right) interface (see Fig.~\ref{fig:setup}). $G_{\rm NL}$ is given by
$G_{\rm NL} = \partial I_{\rm R}/\partial V_{\rm L}$,
where $V_{\rm L}$ is the left lead voltage. The current density $\vec{j}$ at a point $\vec{R}$ in the right electrode is $\vec{j} = e \sum_{\sigma} N_\sigma \int \frac{\mathrm{d} \varepsilon}{8 \pi \i} \,  \text{Tr} \langle \tau_3 \vec{v}_{{\rm F}\sigma} \hat{g}_\sigma^K(\vec{v}_{{\rm F}\sigma}, \vec{R}, \eps) \rangle$, where $N_\sigma$ is the normal state spin-up (down) DOS in the right FM. Only the trajectory connecting both contacts contributes to $G_{\rm NL}$, and the relevant Keldysh GF $\hat{g}_\sigma^K$ can be calculated from a generalization of Eq.~\eqref{eq:Eilenberger} and the boundary conditions at the interfaces~\cite{Eschrig2009}.
\begin{figure}
\includegraphics[width = 0.48 \columnwidth]{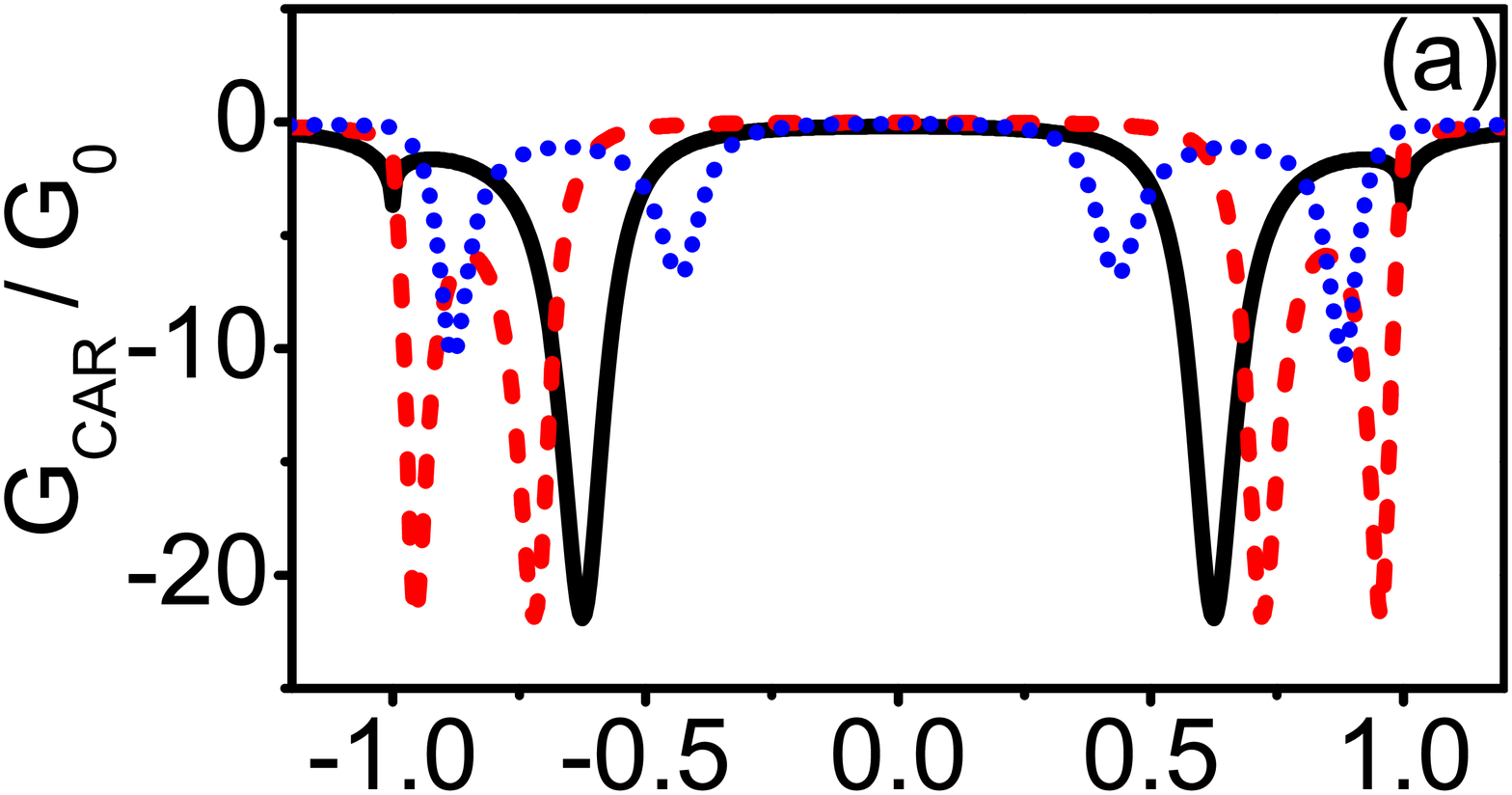}
\includegraphics[width = 0.48 \columnwidth]{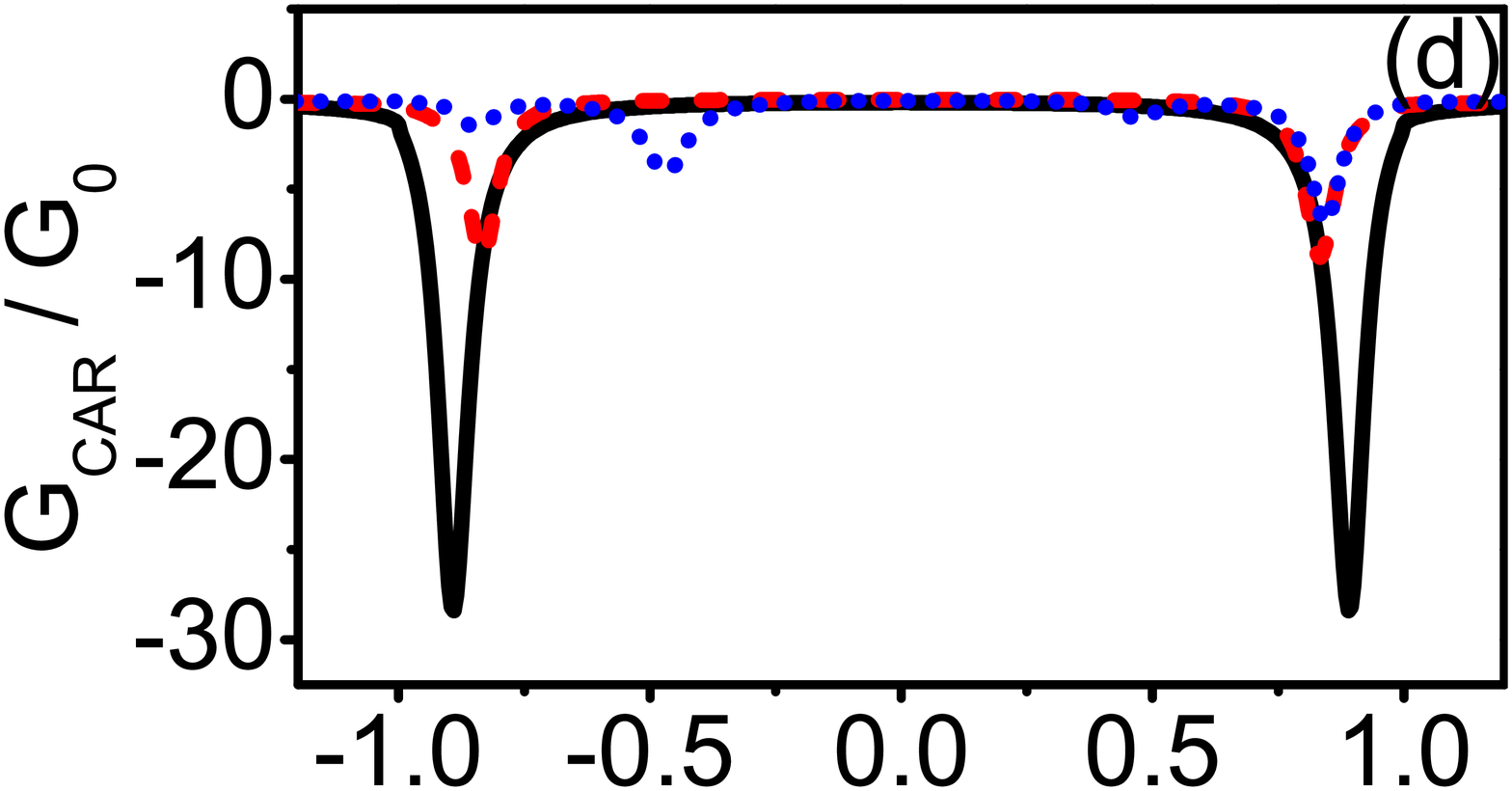} \\
\includegraphics[width = 0.48 \columnwidth]{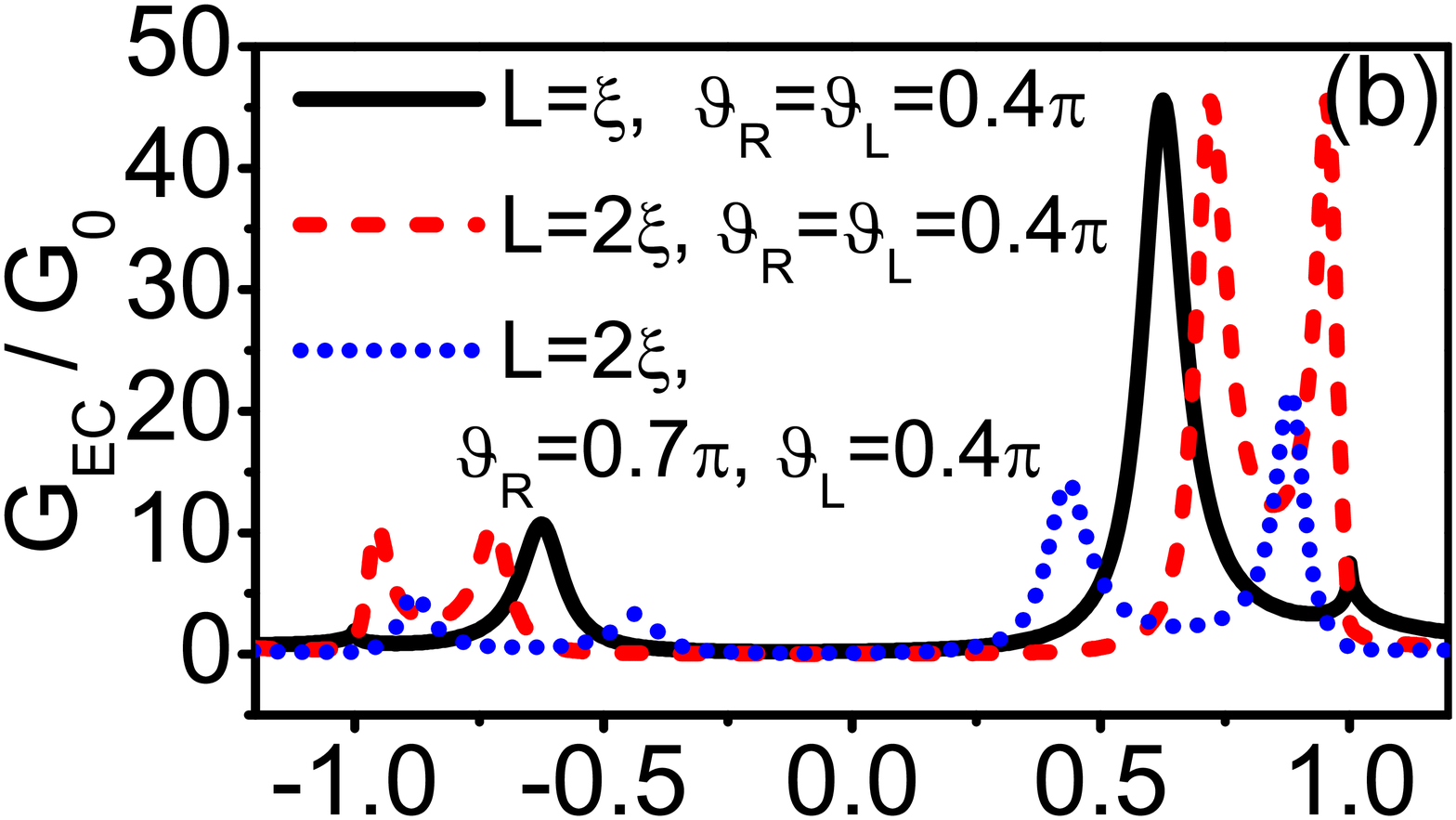}
\includegraphics[width = 0.48 \columnwidth]{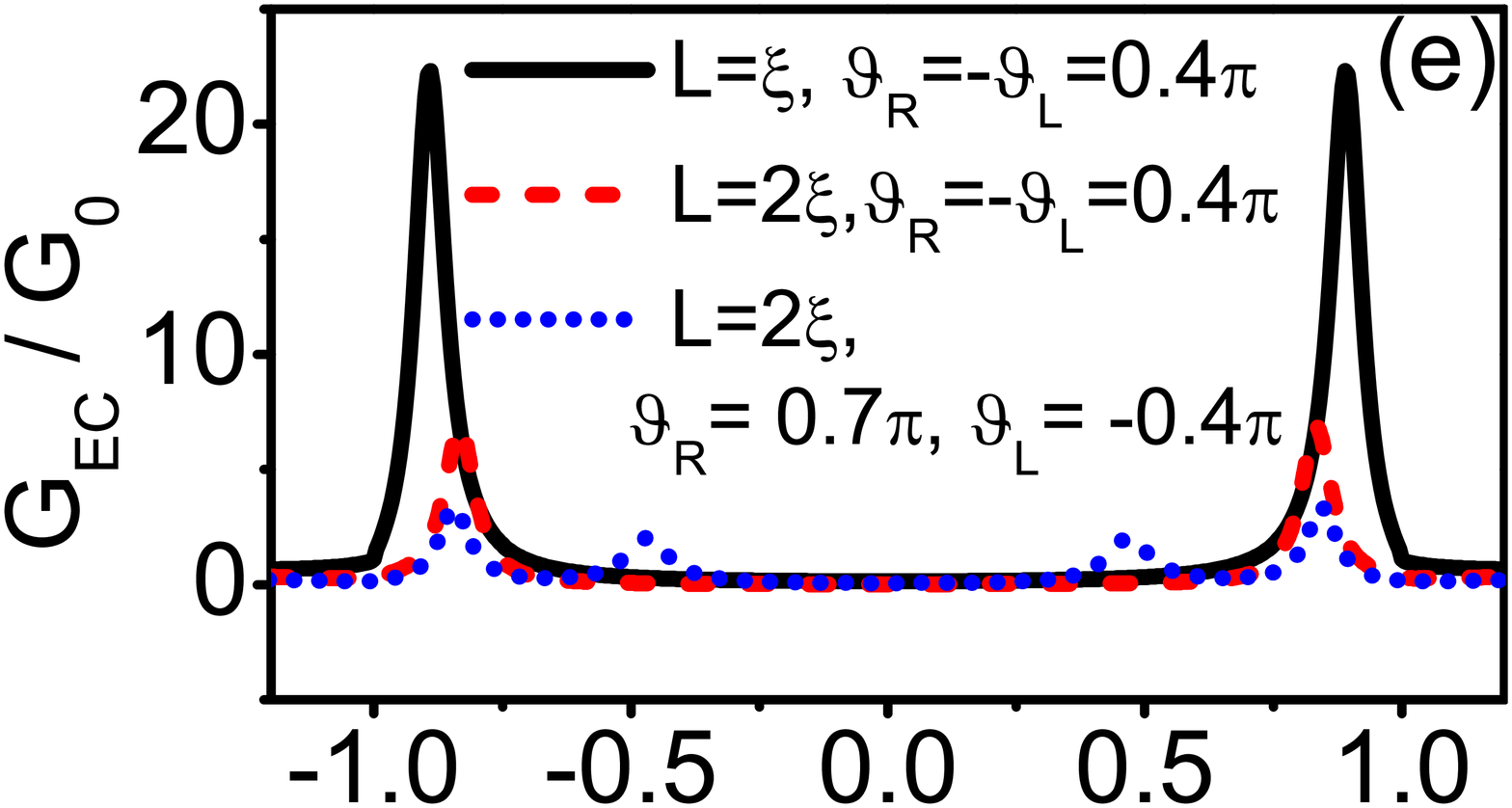} \\
\includegraphics[width = 0.48 \columnwidth]{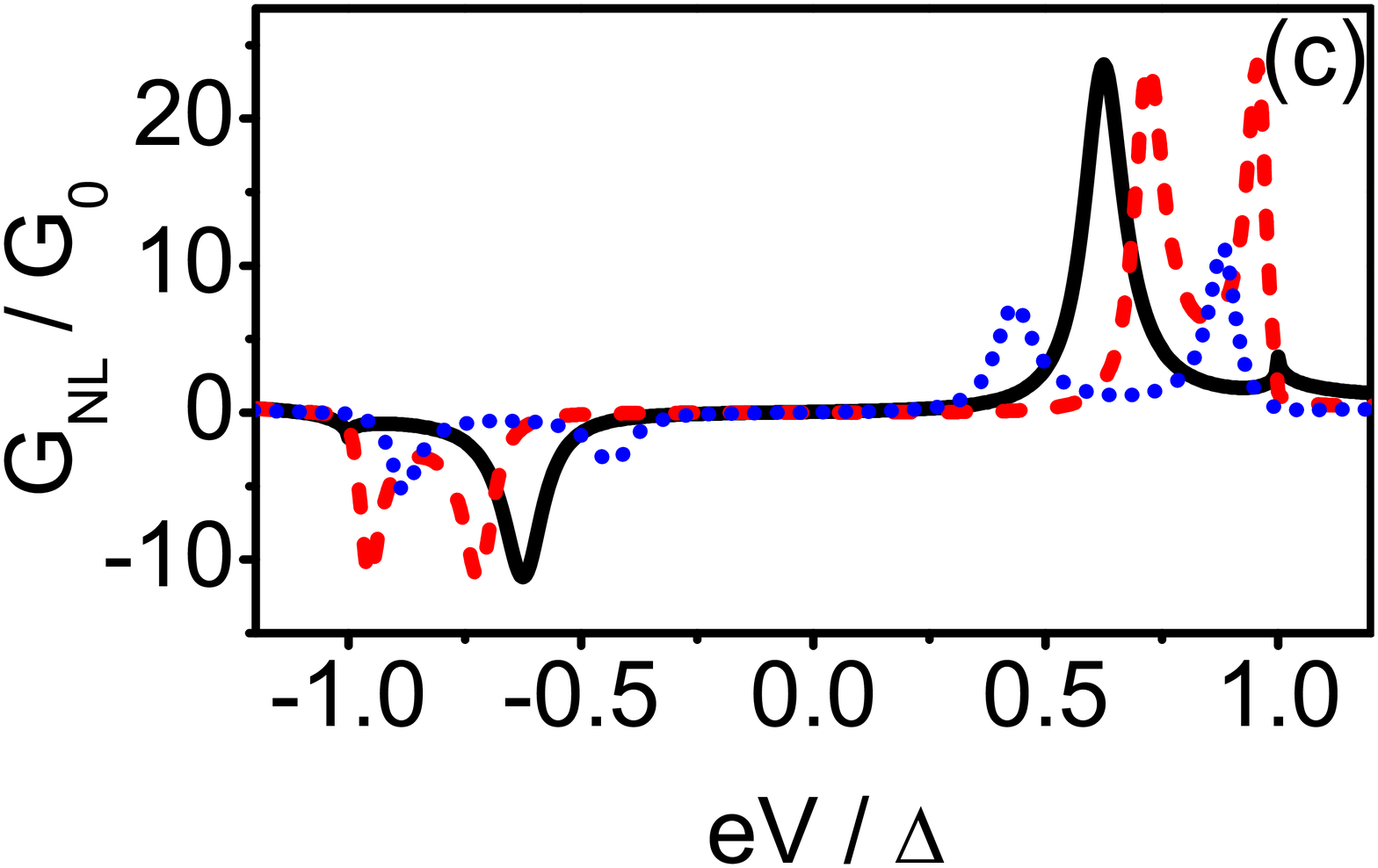}
\includegraphics[width = 0.48 \columnwidth]{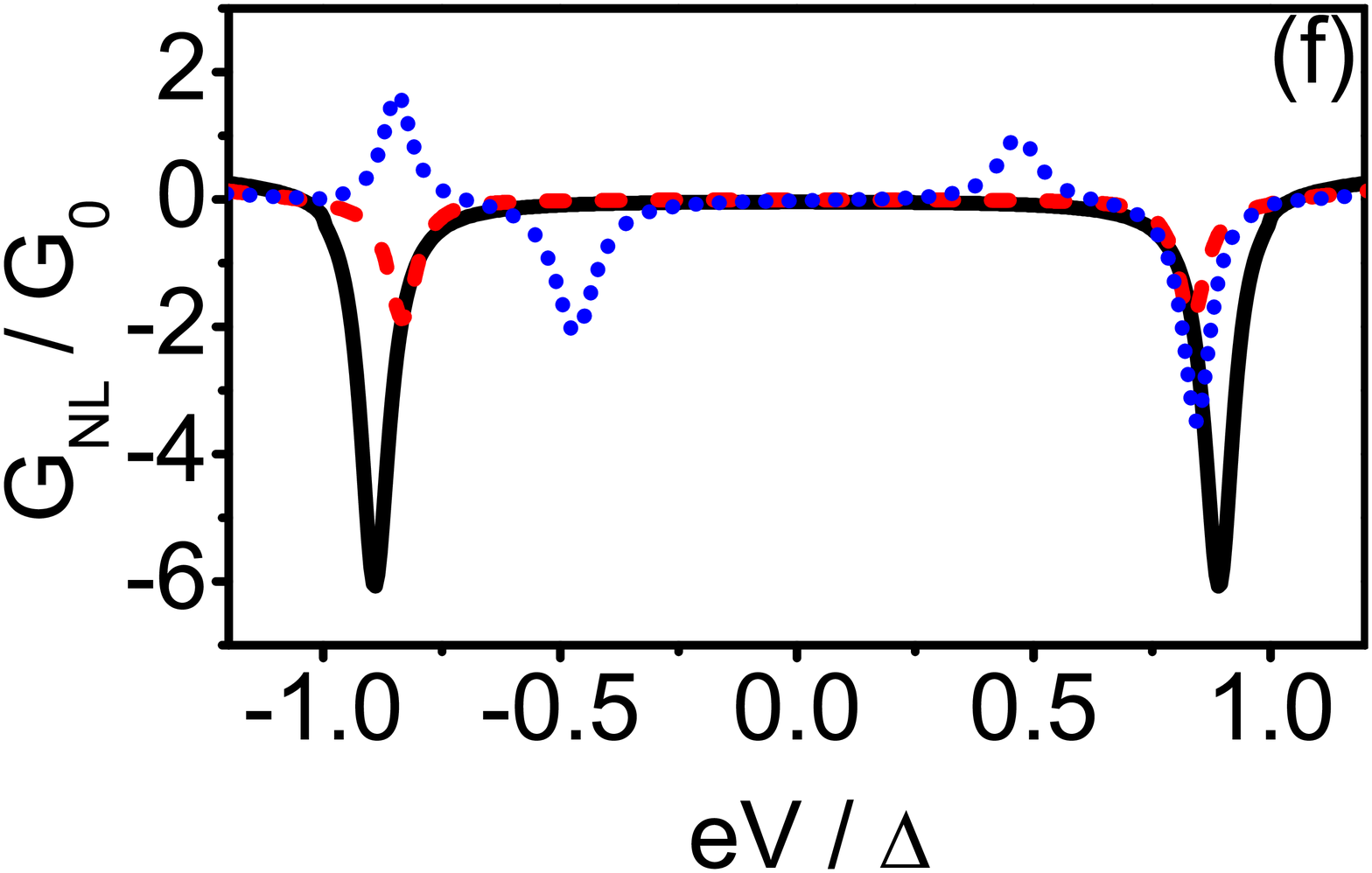}
\caption{(Color online) CAR and EC contributions for P (a-c) and AP (d-f) magnetization of the FM contacts. Transmission coefficients:
(a-c): $t_\sigma^{\rm L} = t_\sigma^{\rm R}$,
(d-f): $t_\sigma^{\rm L} = t_{\bar\sigma}^{\rm R}$, with
$|t_\uparrow^{\rm R}|^2 = 0.14$,
$|t_\downarrow^{\rm R}|^2 = 0.07$.
For other parameters see legends of~(b) and~(e). \label{fig:NL:simple}}
\vspace{-0.5cm}
\end{figure}

\begin{figure}
\begin{minipage}{1.0\columnwidth}
\includegraphics[width = 0.9 \columnwidth]{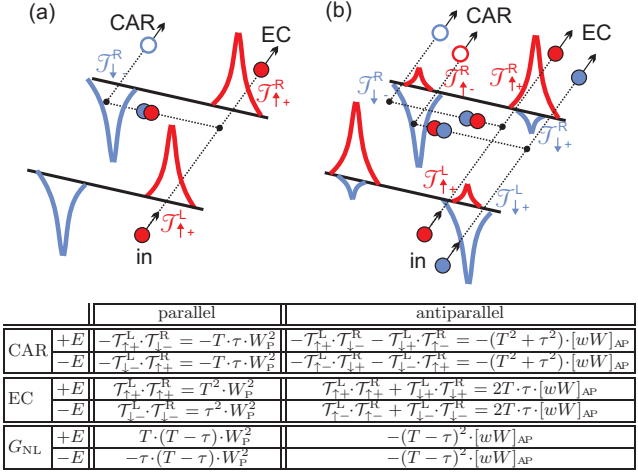}\\[0.2cm]
\resizebox{0.99\columnwidth}{!}{
\begin{tabular}{|l|c||c||c|}
\cline{3-4}
\multicolumn{2}{c||}{} & parallel & antiparallel \\
\hline
\hline
\multirow{2}{*}{CAR} & \multirow{1}{*}{$+E $} & \multirow{1}{*}{$-{\cal T}_{\uparrow+}^{\rm L} \! \! \cdot \! {\cal T}_{\downarrow-}^{\rm R} = -T \! \cdot \! \tau \! \cdot \! W^2_\PP $} & $-{\cal T}_{\uparrow+}^{\rm L} \! \! \cdot \! {\cal T}_{\downarrow-}^{\rm R} - {\cal T}_{\downarrow+}^{\rm L} \! \! \cdot \! {\cal T}_{\uparrow-}^{\rm R} = -(T^2 + \tau^2) \! \cdot \! [w W]_\AP $ \\
\cline{2-4}
& \multirow{1}{*}{$-E$} & \multirow{1}{*}{$-{\cal T}_{\downarrow-}^{\rm L} \! \! \cdot \! {\cal T}_{\uparrow+}^{\rm R} = -T \! \cdot \! \tau \! \cdot \! W^2_\PP $} & $-{\cal T}_{\uparrow-}^{\rm L} \! \! \cdot \! {\cal T}_{\downarrow+}^{\rm R} - {\cal T}_{\downarrow-}^{\rm L} \! \! \cdot \! {\cal T}_{\uparrow+}^{\rm R} = -(T^2 + \tau^2) \! \cdot \! [w W]_\AP $ \\
\hline
\hline
\multirow{2}{*}{EC} & \multirow{1}{*}{$+E$} & \multirow{1}{*}{${\cal T}_{\uparrow+}^{\rm L} \! \! \cdot \! {\cal T}_{\uparrow+}^{\rm R} = T^2 \! \cdot \! W^2_\PP $ } & ${\cal T}_{\uparrow+}^{\rm L} \! \! \cdot \! {\cal T}_{\uparrow+}^{\rm R} + {\cal T}_{\downarrow+}^{\rm L} \! \! \cdot \! {\cal T}_{\downarrow+}^{\rm R} =2 T \! \cdot \! \tau \! \cdot \! [w W]_\AP $ \\
\cline{2-4}
& \multirow{1}{*} {$-E$} & \multirow{1}{*}{${\cal T}_{\downarrow-}^{\rm L} \! \! \cdot \! {\cal T}_{\downarrow-}^{\rm R} = \tau^2 \! \cdot \! W^2_\PP $} & ${\cal T}_{\uparrow-}^{\rm L} \! \! \cdot \! {\cal T}_{\uparrow-}^{\rm R} + {\cal T}_{\downarrow-}^{\rm L} \! \! \cdot \! {\cal T}_{\downarrow-}^{\rm R} = 2 T \! \cdot \! \tau \! \cdot \! [w W]_\AP $ \\
\hline
\hline
\multirow{2}{*}{$G_{\rm NL}$} & \multirow{1}{*}{$+E$} & \multirow{1}{*}{$ \quad \; T\! \cdot \! (T-\tau) \! \cdot \! W^2_\PP $ } & $ -(T-\tau)^2 \! \cdot \! [w W]_\AP $ \\
\cline{2-4}
& \multirow{1}{*} {$-E$} & \multirow{1}{*}{$ -\tau\! \cdot \! (T-\tau) \! \cdot \! W^2_\PP $} & $-(T-\tau)^2 \! \cdot \! [w  W]_\AP $ \\
\hline
\end{tabular}
}
\end{minipage}
\caption{(Color online) Sketch of the spin-up (positive axis) and spin-down (negative axis) Andreev states at the FM/SC interfaces for P (a) and AP (b) orientation of the contact magnetizations. CAR and EC processes are shown, and expressions for obtaining their relative size are given in the table. \label{fig:NLexplained}}
\vspace{-0.5cm}
\end{figure}

The clearest ABS signature occurs for high spin-mixing angles (ABS deep inside the gap), and for the tunneling limit (ABS narrow and well-defined). Fig.~\ref{fig:NL:simple} shows $G_{\rm NL}$ for such a set of parameters. A peaked structure is visible, hinting at the subgap states, with a marked voltage asymmetry in the case of P magnetization. These observations will now be explained with the help of Fig.~\ref{fig:NLexplained}.

For voltages $eV_{\rm L}$ that align the Fermi energy in the left FM with a spin-resolved ABS energy $\varepsilon_{\rm b}$ in the SC, a quasiparticle with the corresponding spin can tunnel from the FM into the SC. In case of EC, the particle has to tunnel to the same spin band at the same energy in the right electrode, while for CAR, an electron with energy $-\varepsilon_{\rm b}$ in the opposite spin band at the right interface is absorbed by the SC to form a Cooper pair (see Fig.~\ref{fig:NLexplained}). Both CAR and EC thus involve (to lowest order) two transmission processes, and the magnitude of the corresponding conductance peak is given by ${\cal T}_{\sigma\pm}^{\rm L} \cdot {\cal T}_{\sigma\pm}^{\rm R}$ for EC and ${\cal T}_{\sigma\pm}^{\rm L} \cdot {\cal T}_{\bar{\sigma}\mp}^{\rm R}$ for CAR, where the ${\cal T}$ are effective transmission coefficients involving the product of the relevant spin-resolved transmission probability and ABS peak height at the left and right interface. The subscript $\pm$ corresponds to the sign of $\varepsilon_{\rm b}$. The relative height of CAR and EC peaks in different magnetization geometries can be fully understood with this simple picture.

As an example, we consider $|t_{\uparrow(\downarrow)}^{\rm L}|^2 = |t_\uparrow^{\rm R}|^2 \equiv T$ and $|t_{\downarrow(\uparrow)}^{\rm L}|^2 = |t_\downarrow^{\rm R}|^2 \equiv \tau$ with $\tau<T\ll 1$ in the P (AP) configuration sketched in Fig.~\ref{fig:NLexplained}. In the P case, there is a single spin-resolved state at energies $\pm \varepsilon_{\rm b}$, and all states have the same weight $W_\PP $ at the interfaces. The EC conductance peak thus scales as $T^2 W^2_\PP $ ($\tau^2 W^2_\PP $) for positive (negative) energies, and the voltage asymmetry in the EC signal in Fig.~\ref{fig:NL:simple}(a) is due to the different transmissions for spin-up and spin-down. The CAR conductance scales as $-T \, \tau \, W^2_\PP $, both for $eV_{\rm L} = \pm \varepsilon_{\rm b}$, and is therefore symmetric in $eV_{\rm L}$. As a result, the total NL conductance in Fig.~\ref{fig:NL:simple}(c) changes sign so that one could switch between CAR or EC by tuning $eV_{\rm L}$. The splitting of the conductance peaks for longer lengths [dashed curves in Figs.~\ref{fig:NL:simple}(a-c)] is due to the ABS repulsion discussed in Fig.~\ref{fig:LDOSvsLengthandtheta}(a). In the AP case, both a spin-up and spin-down state are present at every ABS energy so both CAR and EC contain the sum of two contributions: one from each spin-state. For EC the conductance peak scales as $2 \, T \, \tau \, [w W]_\AP $, while the CAR signal scales as $-(T^2 + \tau^2) \, [w W]_\AP $, where $w_\AP $ ($W_\AP $) is the height of the smaller (larger) ABS peak (for Fig.~\ref{fig:NL:simple}, solid lines, $[wW]_\AP  =0.45 W^2_\PP $). These expressions are valid at $eV_{\rm L} = \pm \varepsilon_{\rm b}$, so the NL conductance signals for $\vartheta_{\rm R}=-\vartheta_{\rm L}$ in Figs.~\ref{fig:NL:simple}(d-f) are all symmetric in voltage.
In contrast to the P case, for the AP case
the conductance peaks are reduced when the distance between the contacts becomes larger (compare solid/dashed curves in Fig.~\ref{fig:NL:simple}) because the weight $w_{\rm AP}$ decreases with the interface distance.

In more general cases, where $|\vartheta_{\rm L}| \neq |\vartheta_{\rm R}|$ as in the dotted curves in Fig.~\ref{fig:NL:simple}, the NL conductance contributions show peaks at four ABS energies (compare to Fig.~\ref{fig:LDOSvsx}).
Their relative weight can be determined analogously to the considerations above.
E.g., for Fig.~\ref{fig:LDOSvsx}(c-d) there are bound states at
$\pm\varepsilon_{\rm b1} $ and $\pm \varepsilon_{\rm b2}$
(with $|\varepsilon_{\rm b1} |>|\varepsilon_{\rm b2}|$).
For the P case, similar expressions as in Fig.~\ref{fig:NLexplained} hold, with
$W^2_\PP $ replaced by $[w_{\ell} W_{\ell}]_\PP $, where $\ell \in \{1,2\}$ labels the bound states.
In the AP case, the EC contribution at $\pm \varepsilon_{{\rm b}\ell} $ is $T \tau [w_{\ell} W_{\ell}]_\AP $, and
the CAR contributions are asymmetric in voltage: for
$eV_{\rm L}=\{-\varepsilon_{\rm b1},-\varepsilon_{\rm b2},\varepsilon_{\rm b2},\varepsilon_{\rm b1} \}$ they are
$\{ -\tau^2 , -T^2 \gamma, -\tau^2 \gamma, -T^2 \}  [w_{1} W_{1}]_\AP $, with $\gamma = \left[ w_{2} W_{2} / w_{1} W_1 \right]_\AP $~\cite{footnote}. The corresponding
relative contributions to $G_{\rm NL}$ are
$\{ \tau , -T \gamma, \tau \gamma, -T\} (T-\tau) [w_1 W_1]_\AP$.

In conclusion, we have studied subgap Andreev states in a FM/SC/FM setup. Due to an interaction between the bound states induced at the two SC/FM interfaces, their energetic positions depend on the relative interface magnetization orientation. This leads to marked asymmetries of the non-local conductance, which we explain in terms of CAR and EC processes with a simple picture based on the Andreev state positions and their weight, thereby clarifying the important role of Andreev states in non-local conductance experiments.

\end{document}